\def\cm2{cm$^{-2}$}

\def\c2{C~{\sc ii}}
\def\c4{C~{\sc iv}}
\def\fe2{Fe~{\sc ii}}
\def\fe3{Fe~{\sc iii}}
\def\mg1{Mg~{\sc i}}
\def\mg2{Mg~{\sc ii}}
\def\si2{Si~{\sc ii}}
\def\si4{Si~{\sc iv}}
\def\al2{Al~{\sc ii}}
\def\al3{Al~{\sc iii}}
\def\o1{O~{\sc i}}
\def\n1{N~{\sc i}}
\def\h1{H~{\sc i}}

\def\approxlt{\mathrel{\spose{\lower 3pt\hbox{$\sim$}}
        \raise 2.0pt\hbox{$<$}}}
\def\approxgt{\mathrel{\spose{\lower 3pt\hbox{$\sim$}}
        \raise 2.0pt\hbox{$>$}}}

\documentclass[article]{gwp80}
\usepackage{graphicx}
\usepackage{amssymb}
\tabletypesize{\normalsize}  

\def\plotone#1{\centering \leavevmode
\includegraphics[width=.95\columnwidth]{#1}}

\def\plotone#1{\centering \leavevmode
\includegraphics[width=.95\columnwidth]{#1}}

\shortauthors{Smith, Catelan, \& Kuehn}
\shorttitle{RR Lyrae period-amplitude diagrams}

\begin{document}
\large    
\pagenumbering{arabic}
\setcounter{page}{1}

\title{RR Lyrae Period-Amplitude Diagrams:\\
 From Bailey to Today}

%
%
\author{{\noindent Horace A. Smith {$^{\rm 1}$}, M\'arcio Catelan {$^{\rm 
2}$}, and Charles Kuehn {$^{\rm 1}$}\\
\\
{\it (1) Department of Physics and Astronomy, Michigan State University, East
Lansing, MI, 48824, USA\\
(2) Pontificia Universidad Cat$\rm{\acute{o}}$lica de Chile, Departamento de 
Astronomi\'ia y Astrof\'isica, Santiago, Chile} 
}
}

%
%
\email{(1) smith@pa.msu.edu}
\email{(2) mcatelan@astro.puc.cl}
\email{(3) kuehncha@msu.edu }


\begin{abstract}
More than a century ago, Solon Bailey's pioneering investigations of the 
variable stars in globular clusters allowed the first period-amplitude 
diagrams to be constructed for their RR Lyrae stars.  These diagrams differ 
from cluster to cluster, and there has been debate as to whether these 
differences are correlated mainly with [Fe/H] or with Oosterhoff type.  It is 
clear now that a cluster's Oosterhoff type plays an important role in 
determining its period-amplitude relation, although the Oosterhoff dichotomy 
itself is correlated with metallicity.  Not all clusters follow the usual 
patterns, however.  The globular clusters NGC~6388 and NGC~6441 have period-amplitude diagrams similar to those of metal-poor Oosterhoff type II globular 
clusters, but they themselves are comparatively metal-rich.  The 
period-amplitude diagrams of Oosterhoff-intermediate systems are discussed.

\end{abstract}

\section{Introduction}
The study of the RR Lyrae period-amplitude diagram, as with much else 
concerning RR Lyrae stars
in globular clusters, commences with the work of Solon I. Bailey and his 
collaborators, who pioneered
the photographic investigation of globular cluster variables. Indeed, a plot 
of period versus
amplitude for RR Lyrae stars is now sometimes termed a Bailey diagram, 
although that name does not
seem to have been applied to such diagrams until the 1990s.

The a, b, and c Bailey types for RR Lyrae stars were introduced in his study 
of variable stars in the
globular cluster $\omega$ Centauri (Bailey 1902). These are now usually 
condensed to just two types,
the RRab stars that pulsate in the fundamental radial mode and the RRc stars 
that pulsate in the
first overtone radial mode.  More recently, RRab stars have been termed RR0 
stars
and RRc variables have been termed RR1 stars (Alcock et al. 2000a).  

The close 
association of the
Bailey a and b types is apparent in the period-amplitude diagrams that can be 
constructed even from the
earliest photographic observations.  In Figure~1, we plot the period versus 
the blue photographic 
amplitude of RR Lyrae stars in the globular cluster M15 based upon data from 
Bailey et al. (1919).  There is a clear
distinction between the Bailey c-type and the Bailey a- and b- type variables, 
but the latter form only a single
sequence, though with considerable scatter.  Later studies have shown that 
while some of this scatter can be
attributed to observational error, some is intrinsic to the stars. One source 
of scatter, but not the only source, is the Blazhko effect.  The amplitudes 
of RR Lyrae variables that undergo
the Blazhko effect change during the secondary Blazhko cycle, complicating 
the
determination of the correct amplitude to use in constructing the 
period-amplitude diagram (Smith 1995).

\begin{figure*}
\centering
\plotone{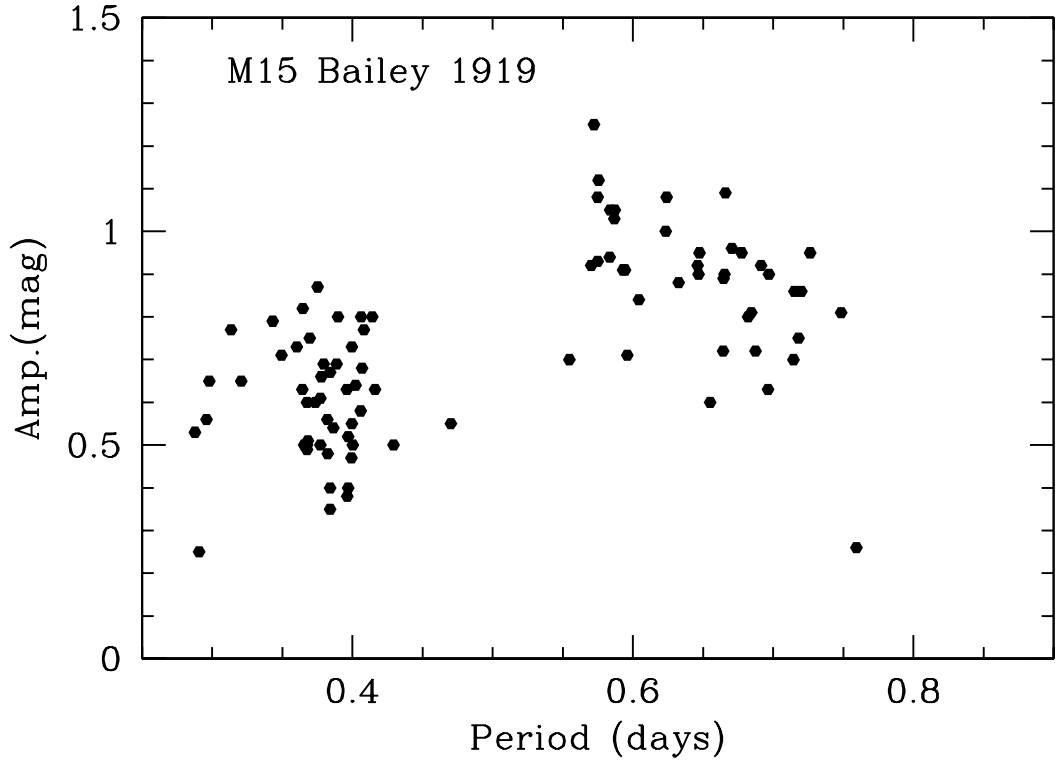}
\vskip0pt
\caption{Period-amplitude diagram for RR Lyrae stars in the globular cluster 
M15, based upon the early work of Bailey et al. (1919).  The RRc stars are 
seen at periods shorter than 0.5 day.  The Bailey a- and b- type variables form 
the group at periods longer than 0.5 day. }
\label{smithfig1}
\end{figure*}

\section{The Period-Amplitude-[Fe/H] relation}

\begin{figure*}
\centering
\plotone{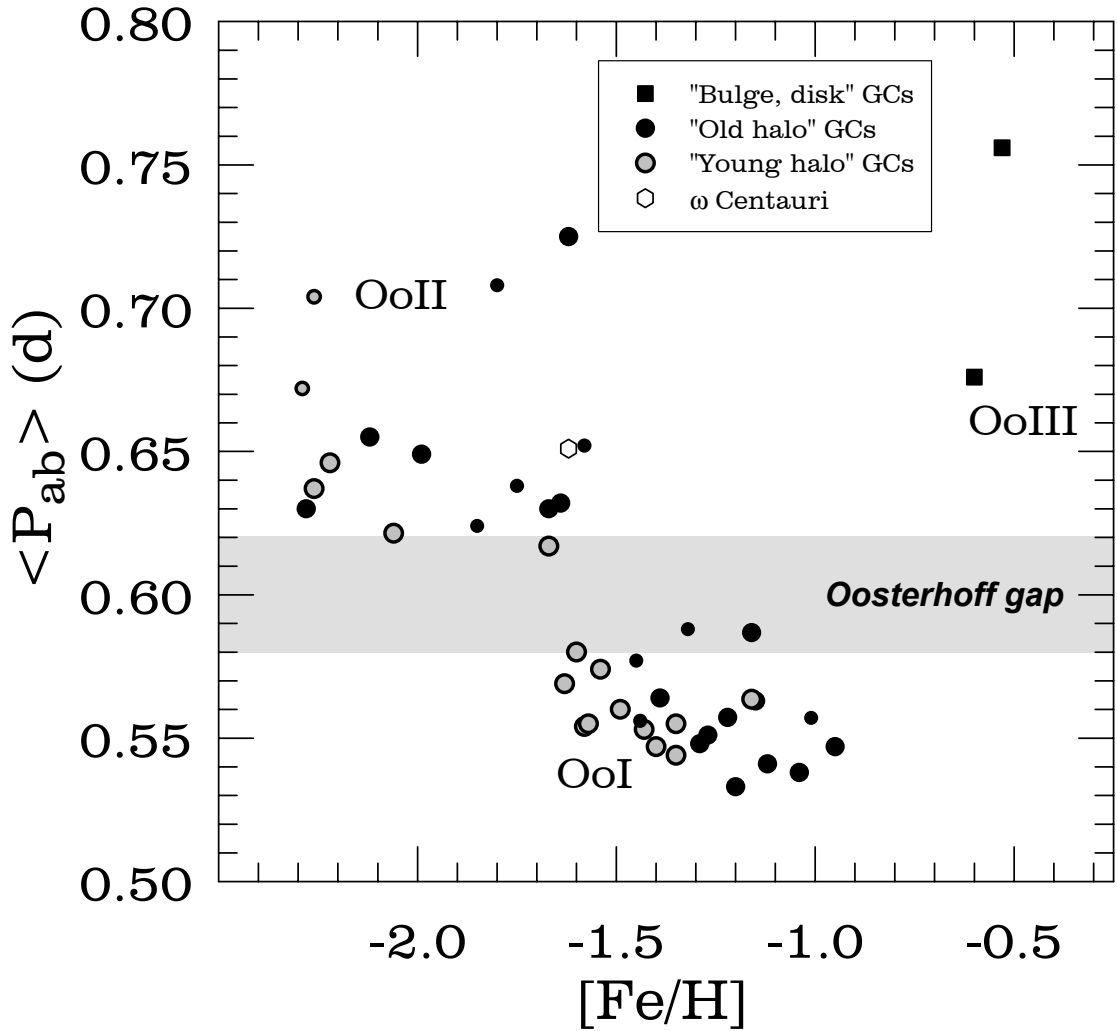}
\vskip0pt
\caption{The Oosterhoff dichotomy is shown in this plot of the mean period of 
RRab stars versus [Fe/H] for Galactic globular clusters.  Also shown are the 
two unusual clusters NGC~6388 and NGC~6441, labeled Oosterhoff III for 
convenience.  These clusters are discussed in 
Section 3. See Catelan (2009) for more details. }
\label{smithfig2}
\end{figure*}

The location of RR Lyrae stars in the period-amplitude diagram is different 
for different globular
clusters, begging the question of what is responsible for the differences.  
One of the most prominent differences is
the distinction between globular clusters of Oosterhoff type I and Oosterhoff 
type II.  RRab stars in Oosterhoff I clusters have mean
periods near 0.55 day, while those in Oosterhoff II clusters have mean
periods near 0.64 day (Oosterhoff 1939; Smith 1995; Clement et al. 2001; Catelan 2009).  Moreover,
Oosterhoff I clusters tend to be less deficient in metal abundance than 
clusters of Oosterhoff II (Figure~2). Preston's (1959) pioneering $\Delta$S 
study
of the metal abundances of field RR Lyrae stars demonstrated that they, too, 
show a decrease in average period with increasing metallicity.

In Figure~3,
we compare the period-amplitude diagrams of RRab stars in the Oosterhoff I 
cluster M3 and the
Oosterhoff II cluster $\omega$ Cen. The $\omega$ Cen variables are shifted to 
a longer period at a given
amplitude than the M3 variables.  In addition, although known Blazhko effect 
stars have been excluded
from this figure, a real scatter among the stars of each cluster remains.

\begin{figure*}
\centering
\plotone{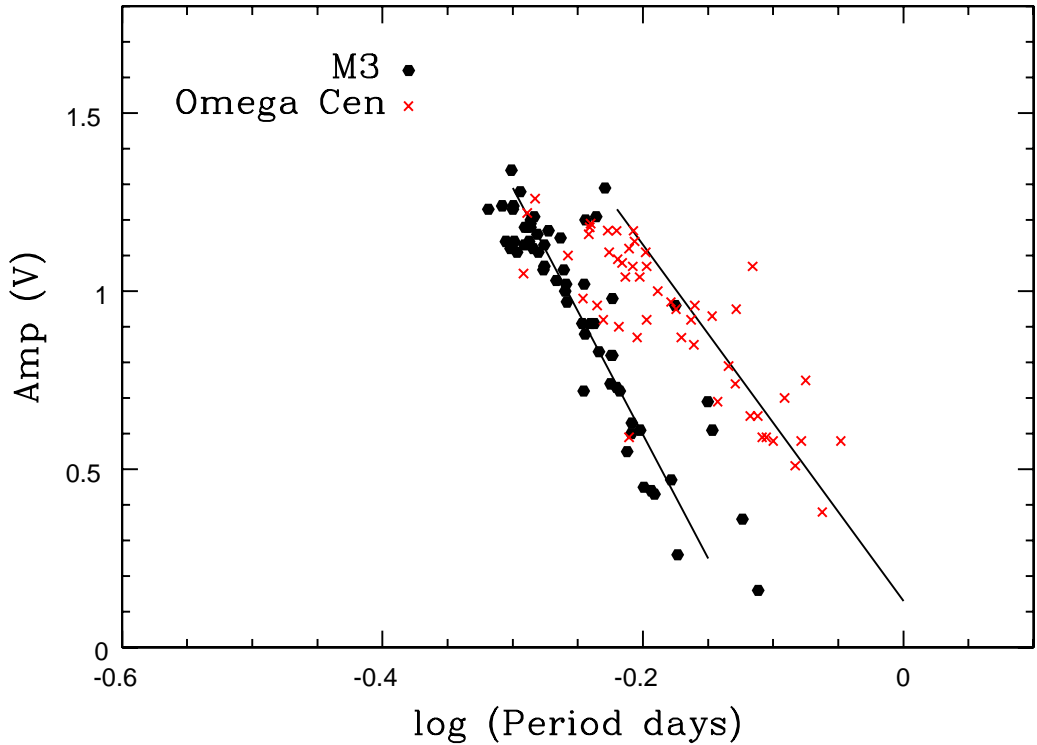}
\vskip0pt
\caption{Period-amplitude relations for RRab stars in the Oosterhoff I 
cluster M3 and the Oosterhoff II cluster $\omega$ Cen, based upon 
Cacciari et al. (2005) and Kaluzny et al. (2004). RR Lyrae stars in
$\omega$ Cen are unusual in having a significant range in [Fe/H] within a 
single cluster.  The solid lines indicate typical locations of the RRab 
period-amplitude relations for Oosterhoff~I and Oosterhoff~II globular 
clusters
(Clement \& Shelton 1999).}
\label{smithfig3}
\end{figure*}

Sandage (1958), employing the pulsation equation P$\sqrt{\rho /\rho_{\sun}}$ = Q,
suggested that the RR Lyrae stars in Oosterhoff~II clusters were about 0.2 
magnitudes
brighter in $V$ than those in Oosterhoff~I clusters. In the early 1980s, 
Sandage extended this
result to general relationships between the location of RR Lyrae stars in the 
period-amplitude
diagram and both absolute magnitude and [Fe/H] (Sandage, Katem, \& Sandage 
1981; Sandage 1982).  In this work, Sandage introduced
the parameter $\Delta$log P, indicating the shift in the period of a star
of a given amplitude relative to a star of the same amplitude in the mean 
period-amplitude relation for the globular cluster M3. Sandage (1982)
obtained the relation $\Delta$log P = 0.116[Fe/H] + 0.173.

Others have more recently adopted versions
of this period-amplitude-[Fe/H] relation to determine metal abundances for 
RRab stars (e.g. Alcock et al. 2000b; Sandage 2004; Kinemuchi et al. 2006). 
These relations imply the existence of a continuous correlation between 
$\Delta$log P and both the luminosity and the metal abundance of an RRab 
star.  However, it was noted that the mean error in the [Fe/H] values 
determined by this method compared to those derived from spectroscopic 
methods could sometimes be large, 0.3 to 0.4 in [Fe/H] (Sandage 2004; 
Kinemuchi et al. 2006).

The robustness of the correlation between the Bailey diagram and [Fe/H] has 
been questioned.  Clement \& Shelton (1999) found that the
$V$ amplitude for a given period was not a function of metal abundance, but 
was instead a function of Oosterhoff type.  This would mean that there was a 
sharp jump in the Bailey diagram at the transition between Oosterhoff I and 
Oosterhoff II clusters.  Bono et al. (2007) also found that the Oosterhoff 
dichotomy rather than [Fe/H] was the critical factor in the determination of 
the period versus amplitude diagram for RRab stars.  Kunder \& Charboyer 
(2009) found that $\Delta$log P for field RRab stars in the Galactic bulge 
did not correlate well with [Fe/H] values determined by other methods.  
Despite these shortcomings, the relatively metal-rich field RRab stars of the 
thick disk and bulge do tend to have periods shorter than those of RRab stars 
in Oosterhoff~I globular clusters at a given amplitude (see, 
for example,
Figure 2 in Kunder \& Charboyer (2009).   Moreover, since the Oosterhoff~I RRab stars have shorter periods than RRab stars in the 
still more metal-poor Oosterhoff~II clusters, a rough correlation 
between the location of the Bailey diagram and [Fe/H] does exist.

The Bailey diagrams of the Oosterhoff I cluster M3 and the Oosterhoff II 
cluster M2 do, however, illustrate the importance of Oosterhoff type.  These 
two globular clusters have similar metallicities.  M3 is at [Fe/H] = -1.57 
and M2 is at [Fe/H] = -1.62, according to Zinn \& West (1984).  Thus, were 
[Fe/H] the main determinant of the location of the cluster RRab stars in the 
Bailey diagram, RR Lyrae stars in M2 and M3 would be expected to fall along 
similar loci in the period-amplitude diagram.  Instead, as shown in Figure~4, 
the M2 stars are shifted to longer periods on average than those in M3.

\begin{figure*}
\centering
\plotone{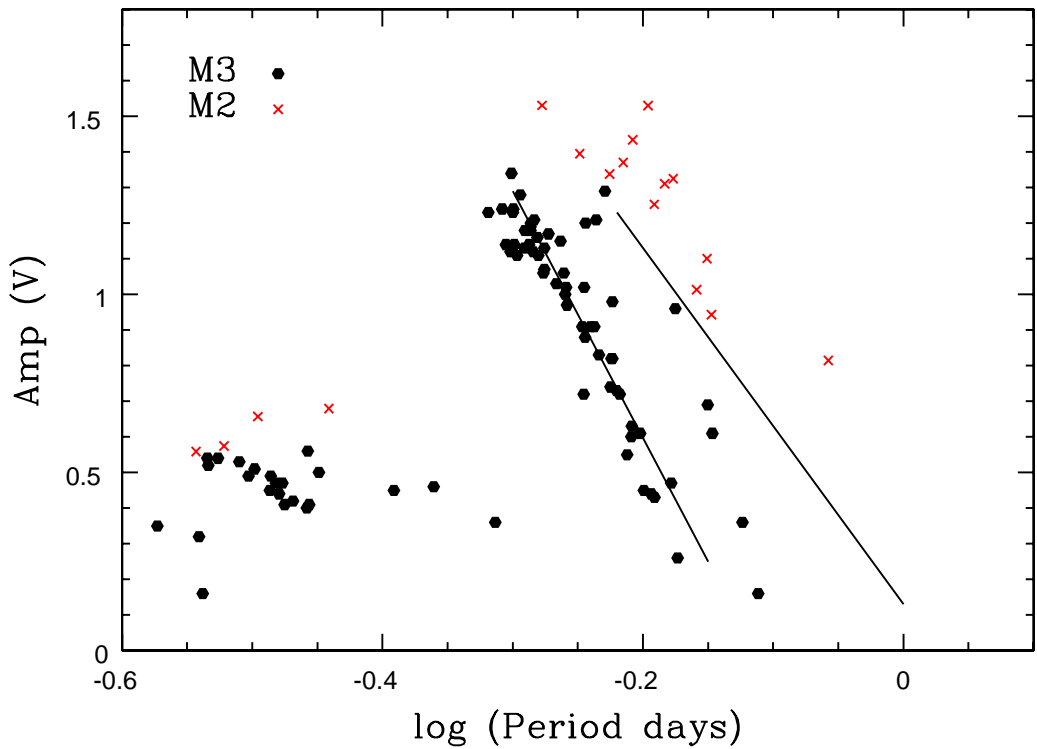}
\vskip0pt
\caption{Period-amplitude diagrams for RR Lyrae stars in the Oosterhoff~I 
cluster M3 (from Cacciari et al. 2005) and the Oosterhoff~II cluster M2 (from 
Lee \& Carney 1999).  These two clusters have similar values of
[Fe/H]. The solid lines are as in Figure~3.
}
\label{smithfig4}
\end{figure*}

It is also noteworthy, however, that the period-amplitude diagrams of RRab stars in the most metal-poor Oosterhoff~II clusters, such as M15 and M68, show considerable scatter, with many stars falling into the region between the mean trends of the two Oosterhoff groups (e.g. figure 4 in Cacciari et al. 2005 and figures 5 and 6 in Corwin et al. 2008). Thus, Oosterhoff type alone does completely describe the location of the RR Lyrae stars in the Bailey diagram.

\vfill\eject

\section{NGC 6388 and NGC 6441: Oddballs}

In Figure~2, two clusters are plotted that have large values of $\langle 
P_{ab} \rangle$ but also relatively high metal abundances.  These are the 
unusual globular clusters NGC~6388 and NGC~6441.  For convenience, these 
clusters have been denoted as Oosterhoff III clusters, to mark their
distinction from the other clusters in Figure~2.  Studies of the RR Lyrae 
stars in these two clusters (Layden et al. 1999; Pritzl et al. 2001, 2002, 
2003; Corwin et al. 2006) show that they have period-amplitude diagrams 
similar to those in Oosterhoff II systems
(Figure~5). Yet the metal abundances of the clusters are relatively high, 
about [Fe/H] = -0.6 (Armandroff \& Zinn 1988; Clementini et al. 2005), 
comparable to the metal abundances of the shorter-period RRab stars of the thick disk and bulge.

\begin{figure*}
\centering
\plotone{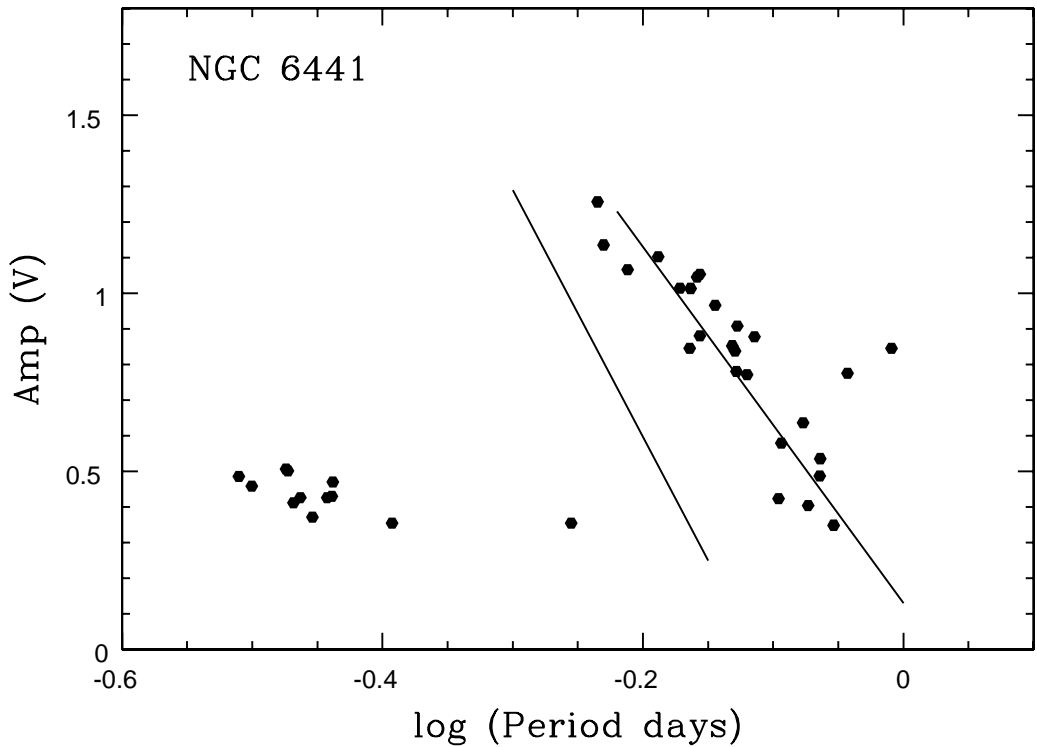}
\vskip0pt
\caption{Period-amplitude diagram for RR Lyrae stars in NGC~6441 (Pritzl
et al. 2001, 2003). The solid lines are the same as in Figure~3.}
\label{smithfig5}
\end{figure*}

Color-magnitude diagrams of NGC 6388 and NGC 6441 (Rich et al. 1997;
Pritzl et al. 2001, 2003; Yoon et al. 2008) show that these two clusters have 
not only the stubby red horizontal branches usually expected in metal-rich 
globular clusters, but also blue horizontal branch components that extend 
through the instability strip.  The long periods of the RR Lyrae stars imply 
that they are more luminous than field RR Lyrae stars of comparable 
metallicity which have shorter periods.  It has been suggested that these 
globular clusters contain more than a single stellar population (Catelan et 
al. 2006; Yoon et al. 2008; Moretti et al. 2009), with the smaller population responsible for the 
blue horizontal branch component possibly being enhanced in helium.  
Though NGC~6388 and NGC~6441 are globular clusters that are far from 
ordinary, they again show that factors other than [Fe/H] can determine the 
location of the RR Lyrae period-amplitude diagram.

\section{Oosterhoff-Intermediate Clusters}

As shown in Figure~2, globular clusters in the Milky Way avoid the Oosterhoff 
gap, the region of $\langle P_{ab} \rangle$ around 0.60 days.  On the other 
hand, globular clusters and dwarf galaxies in the satellite systems of the Milky Way do not show the Oosterhoff 
dichotomy but populate the Osterhoff gap, as shown in Figure~6 (see also 
Catelan 2009; Smith et al. 2009).  There are at least two alternative ways by which 
nature might produce a value of $\langle P_{ab} \rangle$ intermediate between 
the two Oosterhoff groups.  The
intermediate value of $\langle P_{ab} \rangle$ might result from a mixture of 
stars of Oosterhoff types I and II, or the intermediate mean period might 
indicate the existence of a Bailey diagram falling between those of the two 
Oosterhoff groups.  A possible third way of producing an 
Oosterhoff-intermediate cluster is discussed below.

\begin{figure*}
\centering
\plotone{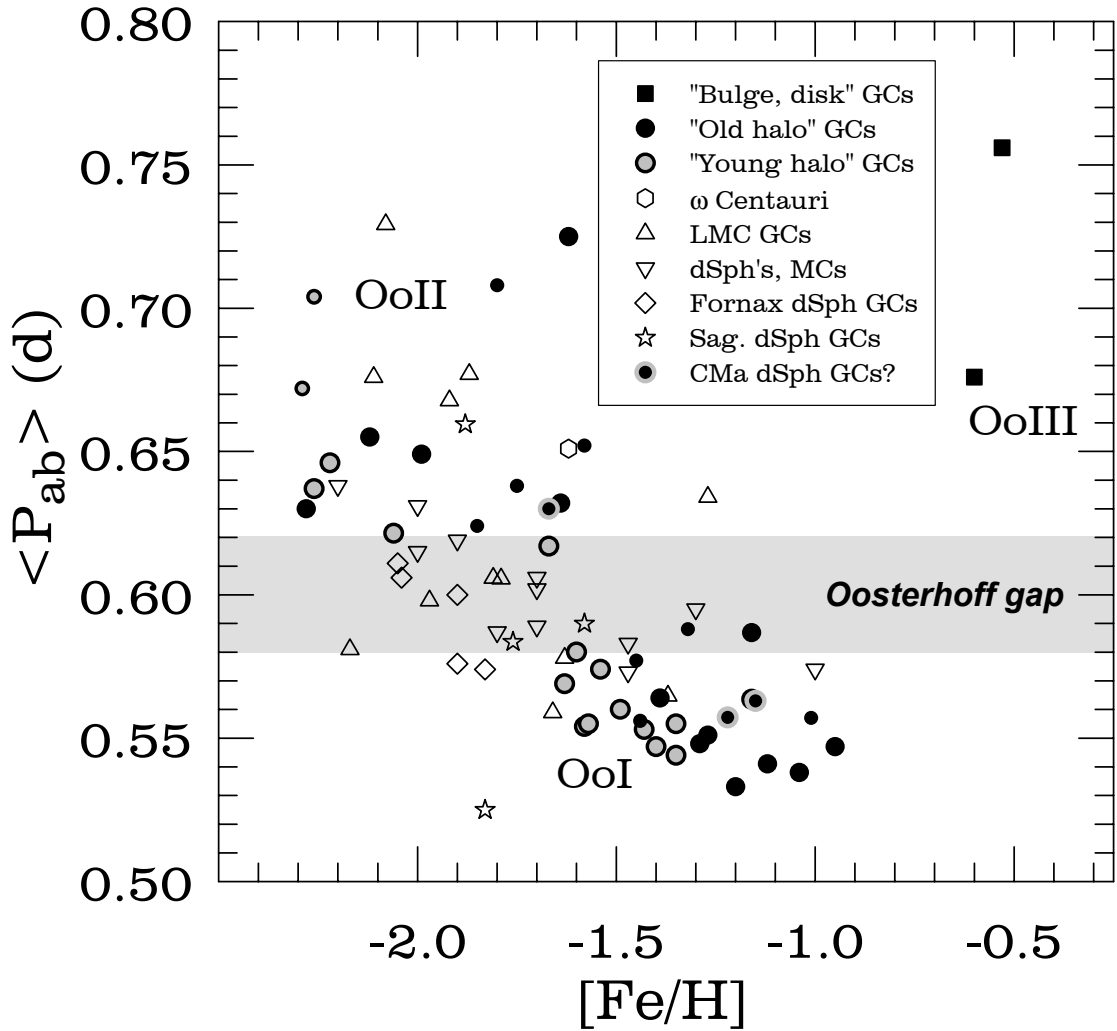}
\vskip0pt
\caption{The Milky Way's dwarf companion galaxies and their globular clusters 
have been added to the Galactic globular clusters shown in Figure~2. The 
Oosterhoff gap is erased by these additions. See Catelan (2009) and Smith et 
al.(2009) for more details. }
\label{smithfig6}
\end{figure*}

Figure~7 shows the Bailey diagram for one of these Oosterhoff-intermediate 
systems, the Draco dwarf spheroidal galaxy (Kinemuchi et al.
2008), which has a value of $\langle P_{ab} \rangle$ = 0.615 days.  It is 
clear from this figure that the intermediate period in this case does not 
arise from an even split of the periods between the two Oosterhoff groups.  
Instead, the individual RRab stars scatter along a mean line that falls 
slightly to the long period side of the Bailey diagram of an Oosterhoff I 
cluster. 

\begin{figure*}
\centering
\plotone{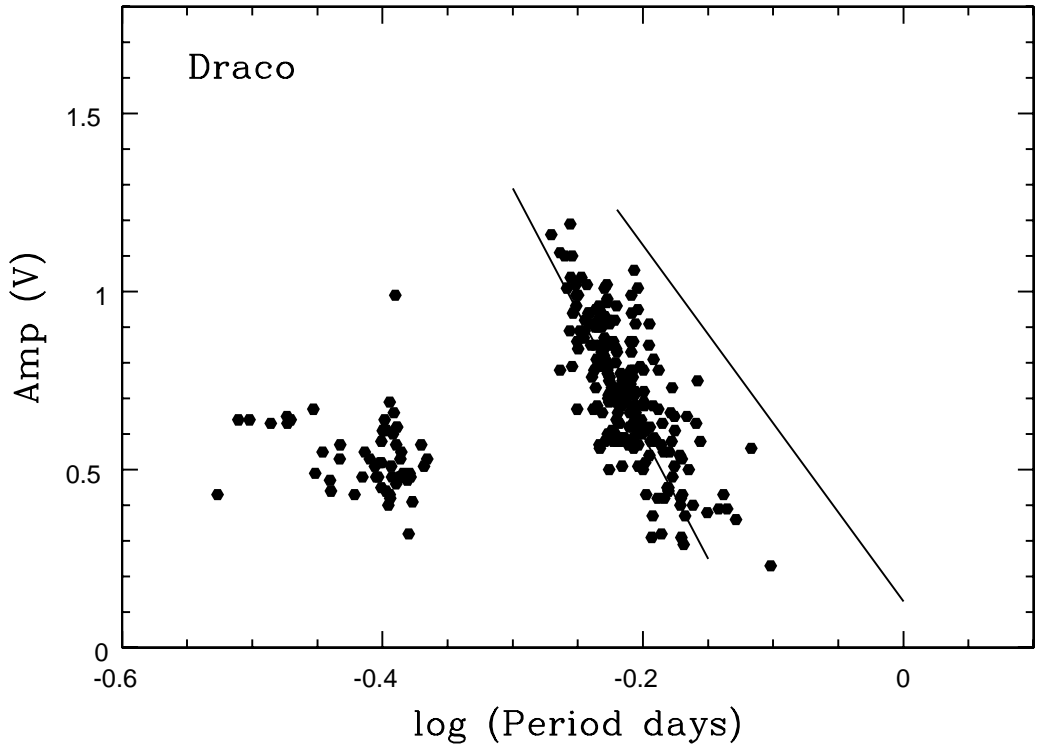}
\vskip0pt
\caption{The Bailey diagram of the Draco dwarf spheroidal galaxy, an 
Oosterhoff-intermediate system (Kinemuchi et al. 2008). The solid lines are 
the same as in Figure~3. Double-mode RRd stars are plotted at the periods of 
their dominant first overtone periods.}
\label{smithfig7}
\end{figure*}

Nonetheless, the Bailey diagrams of all Oosterhoff-intermediate systems are 
not identical.  In Figure~8, we compare the Bailey diagram for the globular 
cluster NGC~1466, located in the outskirts of the Large Magellanic Cloud 
(Kuehn et al. 2011), to that of Draco.  NGC~1466 has a mean period of 
$\langle P_{ab} \rangle$ = 0.59 days, slightly shorter than that of Draco.  
The NGC~1466 RRab stars in Figure~8 fall in a location similar to those in 
Draco.  However, the shortest period RRab star in NGC~1466 has a period 
significantly shorter than the shortest period RRab star in Draco. Draco 
contains many more RRab stars than does NGC~1466, so that this cannot be a 
consequence solely of small-number statistics.

In Figure~8, we also note that the bulk of the RRc stars in NGC~1466 are 
displaced to shorter periods than those in Draco.  In part, this may reflect 
a different distribution of stars along the horizontal branch. However, the 
longest-period, first overtone mode pulsators in Draco (actually double-mode RR Lyrae with a dominant first overtone mode, as noted below) clearly occur at longer 
periods than those in NGC~1466.  Taken together with the shorter period 
cutoff for the RRab stars in NGC~1466, this would be consistent with the 
transition between RRc and RRab stars in NGC~1466 occurring at a shorter 
fundamental mode period than for the Draco RR Lyrae. Note that such a change in the transition period might be a third way of producing clusters with Oosterhoff-intermediate values of $\langle P_{ab} \rangle$.
  
Both NGC~1466 and Draco contain double-mode RR Lyrae stars (RRd stars), 
pulsating simultaneously in the first overtone and fundamental radial modes 
(Kuehn et al. 2011; Nemec 1985; Kinemuchi et al. 2008). When these stars are 
plotted in the Petersen diagram in Figure~9, we see another distinction 
between the two systems.  All except one of the Draco RRd stars have first 
overtone mode periods near 0.55 day and period ratios near 0.746, 
similar to the values usually seen among RRd stars in Oosterhoff type II 
globular clusters (Popielski et al. 2000).  On the other hand, the NGC~1466 RRd stars have first 
overtone periods near 0.48 day and period ratios near 0.744, comparable 
to those seen among Oosterhoff type I globular clusters.  In neither system 
do the RRd stars show properties intermediate between those of RRd stars in 
Oosterhoff I and II systems.

\begin{figure*}
\centering
\plotone{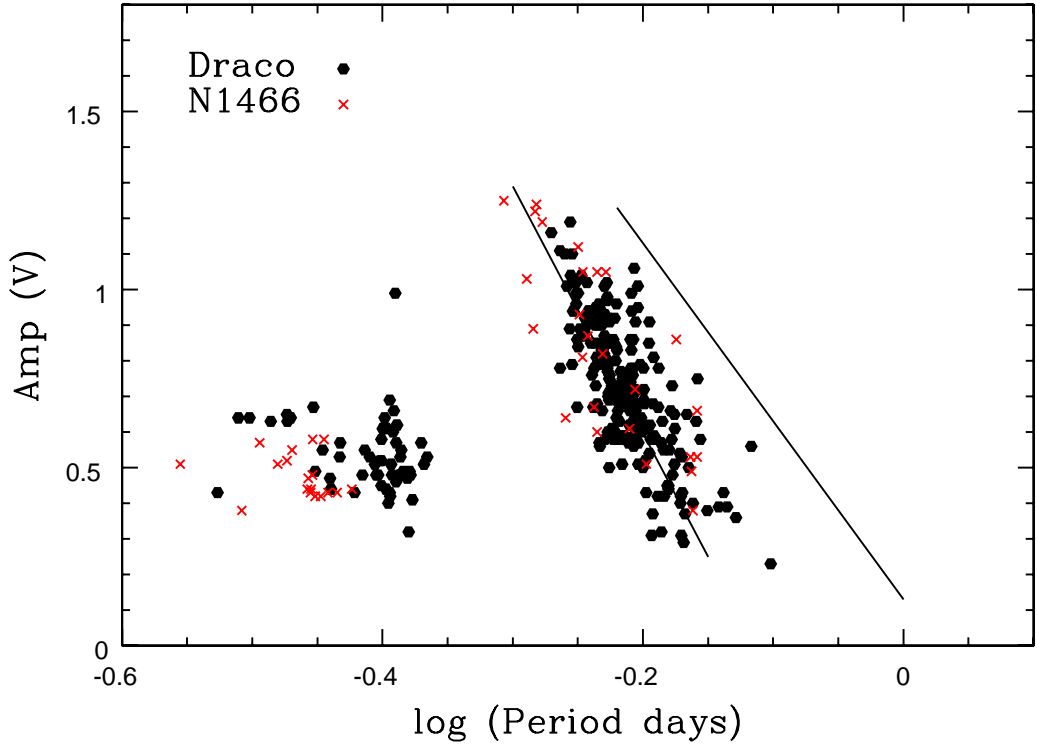}
\vskip0pt
\caption{The Bailey diagram of NGC~1466 is compared to that of the Draco 
dwarf spheroidal galaxy, an Oosterhoff-intermediate system (Kinemuchi et al. 
2008). The solid lines are the same as in Figure~4. Double-mode RRd stars are 
plotted at the periods of their dominant first overtone periods.}
\label{smithfig8}
\end{figure*}

\begin{figure*}
\centering
\plotone{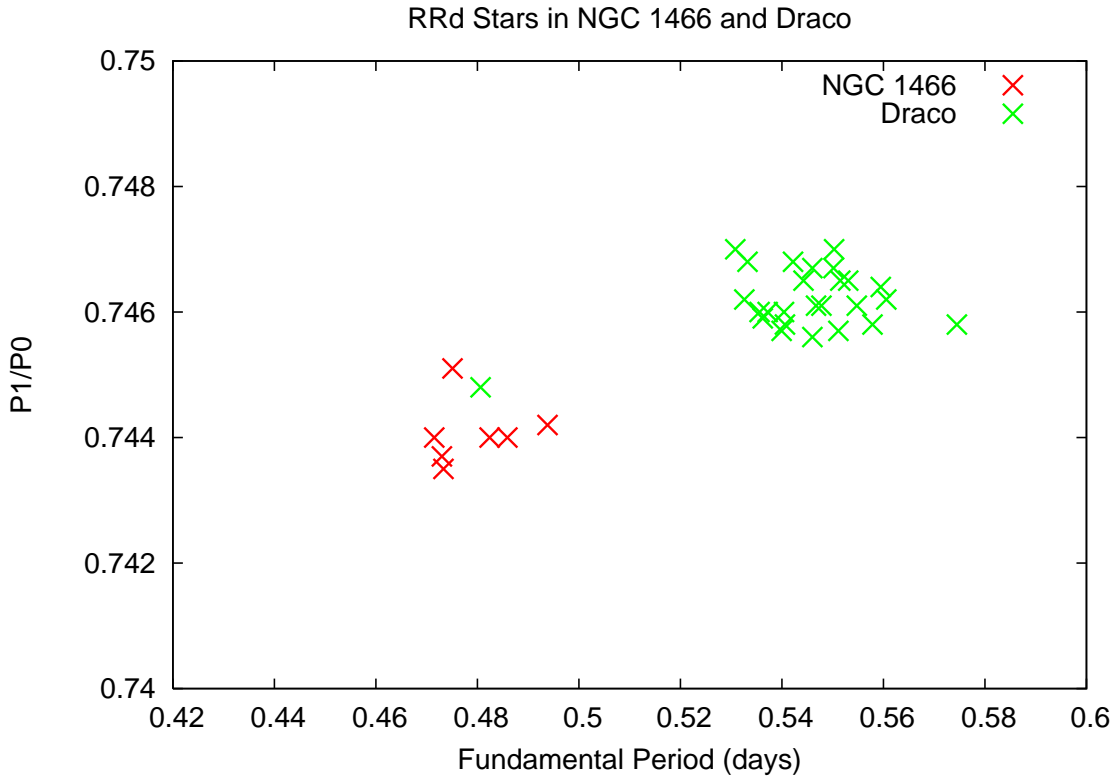}
\vskip0pt
\caption{Double-mode RR Lyrae stars in Draco (Kinemuchi et al. 2008) and NGC~1466 (Kuehn et al. 2011) are plotted in a Petersen diagram of the period ratio of the first overtone to fundamental mode period versus the fundamental mode period.}
\label{smithfig9}
\end{figure*}

\vfill\eject 

\section{Discussion}

While there is generally a rough correlation of the location of the period-amplitude diagram and [Fe/H], there are other factors, particularly 
Oosterhoff type, that determine the location of the period-amplitude 
relation.  Thus, one cannot rely upon period and amplitude to obtain more 
than an approximate value of [Fe/H] for an RR Lyrae star.  The method is, 
however, not entirely useless, and can often give an indication of 
metal abundance that can be valuable in the absence of other information.

The period-amplitude diagrams for RRab stars in the 
Oosterhoff-intermediate 
dwarf galaxies around the Milky Way appear to be genuinely intermediate, 
rather than a mixture of the period-amplitude relations of Oosterhoff~I and 
Oosterhoff~II stars.  However, as the cases of NGC~1466 and Draco illustrate, 
a general similarity in the location of stars in the RRab period-amplitude 
relation does not imply an equal similarity among the RRcd stars.  The 
location of the transition period between the RRab and RRcd stars, as well as 
the distribution of stars in color across the horizontal branch, may play a 
role in producing these differences.

\vskip 1cm

\noindent{\bf Acknowledgements}

Support for M.C. is provided by Proyecto Fondecyt Regular \#1110326; by the 
Ministry for the Economy, Development, and Tourism's Programa Inicativa 
Cient\'{i}fica Milenio through grant P07-021-F, awarded to The Milky Way 
Millennium Nucleus; by Proyecto Basal PFB-06/2007; by FONDAP Centro de 
Astrof\'{i}sica 15010003.  Support for H.A.S. and C.A.K. is provided by NSF 
grants AST 0607249 and AST 0707756.


\begin{thebibliography}

\bibitem{}Alcock, C. et al. 2000a, \apj, 542, 257

\bibitem{}Alcock, C. et al. 2000b, \aj, 119, 2194

\bibitem{} Armandroff, T.~E., \& Zinn, R.\ 1988, \aj, 96, 92


\bibitem{}Bailey, S.~I.\ 1902, Annals of Harvard College Observatory, 38, 1 

\bibitem{} Bailey, S.~I., Leland, E.~F., Woods, I.~E., 
\& Pickering, E.~C.\ 1919, Annals of Harvard College Observatory, 78, 195

\bibitem{}Bono, G., Caputo, F., \& Di Criscienzo, M.\ 2007, \aap, 476, 779 

\bibitem{} Catelan, M., Stetson, P.~B., Pritzl, B.~J., Smith, H.~A., Kinemuchi, K., Layden, A.~C., Sweigart, A.~V., \& Rich, R.~M.\ 2006, \apjl, 651, L133 

\bibitem{} Catelan, M.\ 2009, \apss, 320, 261 

\bibitem{} Cacciari, C., Corwin, T.~M., \& Carney, B.~W.\ 2005, \aj, 129, 267 

\bibitem{} Clement, C.~M., \& Shelton, I.\ 1999, \apjl, 515, L85 

\bibitem{} Clement, C.~M., et al.\ 2001, \aj, 122, 2587 


\bibitem{} Clementini, G., 
Gratton, R.~G., Bragaglia, A., Ripepi, V., Martinez Fiorenzano, A.~F., 
Held, E.~V., \& Carretta, E.\ 2005, \apjl, 630, L145 

\bibitem{} Corwin, T.~M., Sumerel, 
A.~N., Pritzl, B.~J., Smith, H.~A., Catelan, M., Sweigart, A.~V., 
\& Stetson, P.~B.\ 2006, \aj, 132, 1014 

\bibitem{} Corwin, T.~M., 
Borissova, J., Stetson, P.~B., Catelan, M., Smith, H.~A., Kurtev, R., 
\& Stephens, A.~W.\ 2008, \aj, 135, 1459 


\bibitem{} Kaluzny, J., Olech, A., Thompson, I.~B., Pych, W., Krzemi{\'n}ski, W., \& Schwarzenberg-Czerny, A.\ 2004, \aap, 424, 1101 

\bibitem{} Kinemuchi, K., Smith, 
H.~A., Wo{\'z}niak, P.~R., \& McKay, T.~A.\ 2006, \aj, 132, 1202 

\bibitem{} Kinemuchi, K., Harris, H.~C., Smith, H.~A., Silbermann, N.~A., Snyder, L.~A., LaCluyz{\'e}, A.~P., \& Clark, C.~L.\ 2008, \aj, 136, 1921 

\bibitem{}Kuehn, C., et al.\ 2011, \aj, (submitted)

\bibitem{} Kunder, A., \& Chaboyer, B.\ 2009, \aj, 138, 1284 

\bibitem{} Layden, A.~C., Ritter, 
L.~A., Welch, D.~L., \& Webb, T.~M.~A.\ 1999, \aj, 117, 1313 

\bibitem{} Lee, J.-W., \& Carney, B.~W.\ 1999, \aj, 117, 2868

\bibitem{} Moretti, A., et al.\ 2009, \aap, 493, 539 

\bibitem{} Nemec, J.~M.\ 1985, \aj, 90, 204 

\bibitem{} Oosterhoff, P.~T.\ 1939, The Observatory, 62, 104 

\bibitem{} Popielski, B.~L., Dziembowski, W.~A., \& Cassisi, S.\ 2000, AcA, 50, 491 

\bibitem{} Preston, G.~W.\ 1959, \apj, 130, 507

\bibitem{} Pritzl, B.~J., Smith, 
H.~A., Catelan, M., \& Sweigart, A.~V.\ 2001, \aj, 122, 2600 

\bibitem{} Pritzl, B.~J., Smith, 
H.~A., Catelan, M., \& Sweigart, A.~V.\ 2002, \aj, 124, 949 

 \bibitem{} Pritzl, B.~J., Smith, 
H.~A., Stetson, P.~B., Catelan, M., Sweigart, A.~V., Layden, A.~C., 
\& Rich, R.~M.\ 2003, \aj, 126, 1381 

\bibitem{} Rich, R.~M., et al.\ 1997, 
\apjl, 484, L25 


\bibitem{} Sandage, A.\ 1958, Ricerche Astronomiche, 5, 41 

\bibitem{} Sandage, A., Katem, B., \& Sandage, M.\ 1981, \apjs, 46, 41 

\bibitem{} Sandage, A.\ 1982, \apj, 252, 553 

\bibitem{} Sandage, A.\ 2004, \aj, 128, 858 

\bibitem{} Smith, H.~A.\ 1995, Cambridge Astrophysics Series, Cambridge, New York: Cambridge University Press, 1995 

\bibitem{} Smith, H.~A., Catelan, 
M., 
\& Clementini, G.\ 2009, in American Institute of Physics Conference Series, 1170, 179

\bibitem{} Yoon, S.-J., Joo, S.-J., 
Ree, C.~H., Han, S.-I., Kim, D.-G., \& Lee, Y.-W.\ 2008, \apj, 677, 1080 

\bibitem{} Zinn, R., \& West, M.~J.\ 1984, \apjs, 55, 45 




\end{thebibliography}
\end{document}